\documentstyle[prd,aps,eqsecnum]{revtex}
\begin{document}
\title{\bf Free field representation for the O(3) nonlinear sigma \\
model and bootstrap fusion }
\author{ Zal\'an Horv\'ath and G\'abor Tak\'acs \\
         Institute for Theoretical Physics\\
         Roland E\"otv\"os University, Budapest\\
         {\it To appear in Phys. Rev. D} }
\date{19 Oct 1994}
\maketitle
\begin{abstract}
The possibility of the application of the free field representation
developed
by Lukyanov for massive integrable models is investigated in the
context of the O(3) sigma model. We use the bootstrap fusion
procedure
to construct a free field representation for the O(3)
Zamolodchikov-Faddeev
algebra and to write down a representation for the solutions of
the form-factor
equations which is similar to the ones obtained previously
for the sine-Gordon
and SU(2) Thirring models. We discuss also the possibility
of developing
further this representation for the O(3) model and comment
on the extension
to other integrable field theories.
\end{abstract}
\section{Introduction}

Two-dimensional integrable field theories are currently a
prime area of
research in the context of quantum field theory. There is
a huge number of
models in two spacetime dimensions which are exactly
solvable. A lot of
different approaches were developed to deal with these theories.

The first breakthrough in the analysis of integrable models with
massive spectra was the famous paper by
Zamolodchikov and Zamolodchikov
\cite{zamzam}, in
which they calculated the exact S-matrices of some interesting
models using
bootstrap ideas. The
scattering theory in these models is drastically simplified
due to the
existence of an infinite number of conserved currents.
The many-particle
S-matrix is factorized into a product of incoherent
two-particle scattering
amplitudes, hence the name Factorized Scattering Theory (FST).
However one
must realize that in these models
the correspondence between the conjectured exact S-matrices
and the quantum field theory itself is very weak: it is
based on some
information about the spectrum and symmetries of the theory.
There is the
so-called CDD ambiguity which plagues the uniqueness of the
solution for
the S-matrix. In most cases one chooses a "minimal solution"
of the equations
of the FST bootstrap, which means an appropriate fixing of
this ambiguity.

Since the S-matrix is so simple, it doesn't really contain
all information
about the theory: it is an on-shell quantity and for the
calculation of
correlation functions one must go off-shell. However with
the help of the
two-particle S-matrix one can write down equations for
specific matrix elements
of local operators. These are the so-called form-factor
equations and
were used to calculate various off-shell quantities.
With the knowledge of the form-factors
of local fields one can reconstruct all the
quantities associated to
these local operators, such as e.g. correlation functions.
This program was
started long ago \cite{berg}, but the major step forward
was taken by Smirnov,
who in a series of papers calculated the form-factors of some
important operators in the sine-Gordon,
SU(2) Thirring and O(3) nonlinear sigma models
\cite{smirnov1,smirnov2,smirnov3,smirnov4}.

Then a question arises: what is the structure behind the models
which makes
it possible to solve the form-factor equations which is
essentially equivalent
to solving a very
complicated Riemann-Hilbert problem? This line of research was
taken up by
Bernard, Leclair and Smirnov who analysed the nonlocal symmetries
of these
models. They found that the relevant structure is nothing but
quantum affine algebras
and their representation theory \cite{bernard}. Smirnov cleared
up the
connection of the Zamolodchikov-Fateev operators, which play a
very important
role (we discuss them in the sequel) and the quantum affine
symmetry algebra.
It turned out \cite{smirnov5} that the ZF operators are
just vertex operators for the representations of the
quantum affine algebra.
The form-factor axioms proved to be related to the deformed
Knizhnik-Zamolodchikov equations which are the quantum
analogues of the usual
Knizhnik-Zamolodchikov equations for the affine Kac-Moody algebras
\cite{frenkel}. Quantum affine algebras were given a vertex operator
representation in \cite{jing}. This and other conformal field
theory analogies
suggest that a bosonization technique based on free fields
can be useful in
calculating form-factors in the integrable model. This route
was taken up
by Lukyanov \cite{freeaffin,freefield}, who constructed
a free field
representation for the sine-Gordon and SU(2) Thirring models
and used it
to give an integral representation for the form-factors.
In this paper we
use mainly the results of his work \cite{freefield}.

The question we investigate is the following:
how can we apply the bosonization
technique to the O(3) sigma model? Since there is
a simple relationship
(which we call the bootstrap fusion) between
the SU(2) Thirring model S-matrix and the S-matrix
of the O(3) sigma model,
this seems to be rather trivial. However,
it was already noticed by Smirnov, that the corresponding
limit of the SU(2)
Thirring model form-factors is singular and therefore one
must be careful
when trying to define this procedure.
Here we intend to make this fusion procedure
on the level of the free field representation. The other
problem with
this fusion is that the form-factors obtained does not satisfy
one of the
form-factor axioms, namely the one which gives the position
and residues of the
kinematical poles. We identify the source of this problem and
solve it in the
same way as Smirnov did, i.e. by introducing some
multiplier functions which
do not destroy the validity of the other form-factor axioms
but correct
for this problem. The main result is that we obtain a procedure for
calculating generating functions of form-factors in the O(3)
sigma model.

The paper is organized as follows. Section II gives a brief
review of the
Zamolodchikov-Faddeev algebra and an introduction to the free field
representation. Section III defines the notion of the bootstrap
fusion and
discusses the representation of the O(3) ZF-algebra obtained in
this way.
In Section IV we write down the form-factor axioms and give the
idea of the
solution. Then we discuss the regularization of the free field
construction.
Section V contains the result of evaluation of the traces and the
discussion
of the problem of the integration. We then give an explicit example
in section
VI, where we show the calculation of the two-particle form-factor
of the O(3)
current and compare the result to the one found in the literature.
Section
VII is reserved for the discussion. The paper contains an appendix
in which
we show a direct method of proving the ZF-algebra commutation
relations.

\section{Review of the free field representation}\label{contours}

We first study the SU(2) Thirring model described by the Lagrangian
density
\begin{equation}
{\cal L}=i\bar\psi\gamma^\mu\partial_\mu\psi-gj_i^\mu j_{i\mu}.
\label{eq:thlagr}\end{equation}
The field $\psi$ describes an isospin doublet of fermions and the
current
$j_i^\mu$ is the SU(2) current
\begin{equation}
j_i^\mu ={\displaystyle 1\over \displaystyle 2}
\bar\psi\gamma^\mu\tau_i\psi ,
\end{equation}
where the $\tau_i,\ i=1,2,3$ are the Pauli-matrices.
The spectrum of the model consists of a massive isospin doublet of
kinks and
a free scalar. We will be concerned only with the kink sector.
We introduce the
rapidity variable describing the energy and momentum of the
on-shell kink states
by the definition $p^0=m\cosh\beta,\ p^1=m\sinh\beta$. The scattering
processes of the kinks are described by the following S-matrix:
\begin{equation}
S_{i,j}^{k,l}=S_0(\beta )
{\displaystyle\strut 1\over\displaystyle\strut
\beta -\pi i}(\beta\delta_i^k\delta_j^l-\pi i
\delta_i^l\delta_j^k),
\label{eq:thsmat}\end{equation}
where $i,j,k,l=+,-$ denote the isotopic indices and
\begin{equation}
S_0(\beta )={\displaystyle\strut\Gamma({1\over 2}+
{\beta\over 2\pi i})
\displaystyle\strut\Gamma(-{\beta\over 2\pi i})
\over\displaystyle\strut\Gamma({1\over 2}-{\beta\over 2\pi i})
\displaystyle\strut\Gamma({\beta\over 2\pi i})}.
\label{eq:s0}\end{equation}
This S-matrix was obtained using the bootstrap procedure and
the properties
of the conjectured spectrum of the model. These are purely
algebraic reasonings
and the satisfactory connection to the field theory defined by
the Lagrangian
(\ref{eq:thlagr}) is still missing as we discussed in the introduction.
One should also keep in mind however that there are many
cases in which profound arguments can be presented for the
correctness of the
bootstrap approach by comparing with another known approaches such as
perturbation theory, $1/N$ expansion and lattice simulation
(see for example the discussion in \cite{balog}). This gives the
encouragement for the hope that one day we really can establish the
correspondence between the different descriptions.

The charge conjugation matrix for the doublet kinks is $C=i\sigma_2$
which
satisfies the unusual property $C^t=-C$. As it is known in two
dimensions
more general statistics are possible than just bosonic or fermionic
ones.
These new ones correspond to representations of the braid group
instead of the representations of the symmetric group.
Actually the kinks describe
spin-1/4 particles. The S-matrix (and the model) has a global SU(2)
symmetry
under which the kinks transform as the fundamental representation
of the group.

The Hilbert space of an integrable model is defined by
the representation
of the formal Zamolodchikov-Faddeev (ZF) algebra,
which is defined by the
relations
\begin{equation}
Z_i(\beta_1)Z_j(\beta_2)=
S_{i,j}^{k,l}(\beta_{12})Z_l(\beta_2)Z_k(\beta_1),
\qquad \beta_{12}=\beta_1-\beta_2.
\label{eq:scattrel}\end{equation}
The commutation relations of the Z-operators
reflect the scattering of the particles. From now on
we will refer to them
as the scattering relations.
The space of states furnishes a representation of the ZF algebra.
The Hilbert
space structure is fixed by giving the relations for the
adjoint operators
in that representation (we denote it by $\pi_A$ and use the
letter $A$ for
$\pi_A(Z)$)
\begin{eqnarray}
A^{i\dagger}(\beta_1)A^{j\dagger}(\beta_2)&  =
& A^{l\dagger}(\beta_2)A^{k\dagger}(\beta_1)
S^{i,j}_{k,l}(\beta_{12}) \nonumber \\
A_i(\beta_1)A^{j\dagger}(\beta_2)&  =
& A^{k\dagger}(\beta_2)S^{j,l}_{i,k}
(\beta_{12})A_l(\beta_1)+2\pi\delta^j_i\delta(\beta_{12}).
\end{eqnarray}
However in the sequel we will be concerned with another type of
representation
of the ZF algebra in which we do not have the conjugate operators.
Here we would
like to stress that the representation $\pi_A$ is nothing else
but the space of
asymptotic particle states (in and out) of the model as can be
found e.g.
in \cite{freefield}.

These relations can be converted to a new particle basis in which the
fundamental operators of the SU(2) Thirring model are usual spin-1/2
fermions.
To achieve this we redefine the S-matrix (\ref{eq:thsmat}) of the model
as follows
\begin{equation}
\hat S_{i,j}^{k,l}(\beta )=(-1)^{(i+l)}S_{i,j}^{k,l}(\beta)
\end{equation}
where $(-1)^i=+1$ for $i=-$ and $-1$ for $i=+$.
This also means a redefinition
of the phases of the ZF operators. We will drop the hat from now on.
This will
cause no confusion since we will only use the new basis of
auxiliary spin-1/2
particles. The net effect of the phases is that the coproduct
rule for the
SU(2) symmetry changes: we get an SU(2)$\rm_{-1}$ quantum
symmetry as in
\cite{freefield}.
We also give the formula for the S-matrix elements in this
basis in detail:

\begin{eqnarray}
S^{++}_{++}(\beta )&  =& S^{--}_{--}(\beta )=S_0(\beta ), \nonumber \\
S^{+-}_{+-}(\beta )&  =& S^{-+}_{-+}(\beta )=S_0(\beta )
{\beta\over i\pi -\beta}, \nonumber \\
S^{+-}_{-+}(\beta )&  =& S^{-+}_{+-}(\beta )=S_0(\beta )
{ i\pi\over i\pi -\beta}.
\end{eqnarray}

The charge conjugation matrix is given by
\begin{equation}
C_{ij}=\delta_{i+j,0}.
\end{equation}

Now we turn to the free field representation. Lukyanov
\cite{freefield}
introduces a free
field $\phi(\beta )$ with the following properties:
\begin{eqnarray}
& & [\phi(\beta_1),\phi(\beta_2)]=\ln S_0(\beta_2-\beta_1), \nonumber \\
& & \langle 0\vert\phi(\beta_1)\phi(\beta_2)\vert 0\rangle =
-\ln g(\beta_2-\beta_1),
\label{eq:phirelations}\end{eqnarray}
where $S_0$ is the function defined in eqn. (\ref{eq:s0}) and we give
the formula
for the function $g$ below (\ref{eq:gfunct}).
The consistency of the two relations above requires
\begin{equation}
S_0(\beta )={g(-\beta )\over g(\beta )}.
\end{equation}

The field $\phi$ is represented (after a proper ultraviolet
cut-off procedure)
on a Fock space. We will build up operators which represent
the ZF algebra
on that space.

The field
\begin{equation}
\bar\phi(\beta )=
\phi (\beta+i{\pi\over 2})+\phi (\beta-i{\pi\over 2}),
\end{equation}
which satisfies the commutation relations
\begin{equation}
[\bar\phi(\beta_1),\bar\phi(\beta_2)]=
\ln {\beta_2-\beta_1-i\pi\over
\beta_2-\beta_1+i\pi},
\end{equation}
and similar ones with $\phi(\beta )$,
plays an important role as well.
The corresponding two-point functions can also be written down as
\begin{eqnarray}
&  \langle 0\vert\bar\phi(\beta_1)\phi(\beta_2)\vert 0\rangle =
& \ln w(\beta_2-\beta_1), \nonumber \\
&  \langle 0\vert\bar\phi(\beta_1)\bar\phi(\beta_2)\vert 0\rangle =
& -\ln \bar g(\beta_2-\beta_1),
\end{eqnarray}
where we defined the following functions

\begin{eqnarray}
g(\beta )&  =& k^{1\over 2}
{ \Gamma ({1\over 2}+{i\beta\over 2\pi})\over
\Gamma ({i\beta\over 2\pi})}, \nonumber \\
w(\beta )&  =& k^{-1}{ 2\pi\over
 i(\beta +i{\pi\over 2})}, \nonumber \\
\bar g(\beta )&  =& -k^2{\beta(\beta +i\pi )\over
 4\pi^2},
\label{eq:gfunct}\end{eqnarray}
where $k$ is a normalization constant. The field $\bar\phi$ is a
much more
convenient object than the field $\phi$ since it satisfies
much simpler
relations
and its two-point function is closely analogous to the
two-point function
of a free conformal scalar field.

Now we are ready to define the following vertex operators
a la Lukyanov
\begin{eqnarray}
V(\beta )&  =&\exp (i\phi (\beta ))=
(g(0))^{1\over 2}:\exp (i\phi (\beta )):, \nonumber \\
\bar V(\beta )&  =&\exp (-i\bar\phi (\beta ))=
(\bar g(0))^{1\over 2}
:\exp (-i\bar\phi (\beta )):,
\label{eq:vertexdef}\end{eqnarray}
where : denotes an appropriate normal ordering \cite{freefield}.
These operators need to be regularized because the functions $g$
and $\bar g$ have simple zeros at $\beta =0$.
Using the analogy with conformal
theory we define the regularized values of the functions
$g,\bar g$ at
$\beta =0$ to be
\begin{eqnarray}
&  g_{reg}(0)=&\lim\limits_{\beta\rightarrow 0}{ g(\beta )\over
 \beta}=:\rho^2 ,\nonumber \\
&  \bar g_{reg}(0)=&\lim\limits_{\beta\rightarrow 0}
{ \bar g(\beta )
\over \beta}=:\bar \rho^2.
\end{eqnarray}

The vertex operators defined in (\ref{eq:vertexdef}) obey the
following very
important relations
\begin{eqnarray}
&  V(\beta_1)V(\beta_2)=&\rho^2g(\beta_2-\beta_1)
:V(\beta_1)V(\beta_2): \nonumber \\
&  \bar V(\beta_1)V(\beta_2)=&\rho\bar\rho w(\beta_2-\beta_1)
:V(\beta_1)V(\beta_2): \nonumber \\
&  \bar V(\beta_1)\bar V(\beta_2)=&\bar\rho^2
\bar g(\beta_2-\beta_1):\bar V(\beta_1)\bar V(\beta_2):
\label{eq:vertexrel}\end{eqnarray}
Eqn. (\ref{eq:vertexrel}) plays a key role in proving the ZF
scattering relations.

Then in analogy with the Coulomb gas representation of
rational conformal
field theory one introduces a screening charge $\chi$
by the definition
\begin{equation}
\langle u\vert\chi\vert v\rangle = \eta^{-1} \langle u\vert{1\over 2\pi}
\int_C d\gamma\bar V(\gamma )\vert v\rangle ,
\end{equation}
where C is a contour specified in the following manner: assuming
that all matrix elements
$\langle u\vert\bar V(\gamma )\vert v\rangle$
are meromorphic functions decreasing at the infinity faster than
$\gamma^{-1}$,
the contour goes from $\Re \gamma =-\infty$ to
$\Re \gamma =+\infty$
and lies above all singularities whose positions depend on
$\vert u\rangle$
and below all singularities depending on
$\vert v\rangle$. ($\eta$ is a
normalization parameter which is chosen to be $(-i\pi )^{1/2}$. This
normalization guarantees the correct value for the residue of the
operator product (see proposition (ii) below).

Having defined these objects, one can write down the following
operators
\begin{eqnarray}
&  Z_{+}(\beta )=& V(\beta ), \nonumber \\
&  Z_{-}(\beta )=& i(\chi V(\beta )+V(\beta )\chi ).
\end{eqnarray}

Then Lukyanov proves the following propositions:

(i) These operators satisfy the SU(2) Thirring model ZF relations.

(ii) The singular part of the
operator product $Z_i(\beta_2)Z_j(\beta_1)$
considered as a function of the complex variable $\beta_2$
for real $\beta_1$ in the upper half plane
$\Im \beta_2\geq 0$ contains only
one simple pole at $\beta_2=\beta_1+i\pi$ with residue $-iC_{ij}$.

(i) and (ii) can be proven by technics similar to those used in
conformal
field theory. Here we see that the operator $\chi$ plays
the role of the
step operator of the quantum group. It can be compared
with the work of
Gomez and Sierra \cite{gomez}
on the Coulomb gas representation of rational conformal
field theories where they showed that the
positive step operators of the
quantum group can be identified with a contour
creation operator defined
by the help of the screening charges. Here we see a
very close structure
in the definition of the operator $\chi$ and in the formulae for the
bosonization of the ZF operators.

\section{Bootstrap fusion and the ZF algebra of the
nonlinear sigma model}\label{ZFsing}

Here we recall some basic facts about the nonlinear
sigma model. It is defined
to be the dynamics of a scalar field taking values
on the surface of the
two-dimensional sphere $S^2$:
\begin{eqnarray}
&& {\cal L}=
{1\over g^2}\int d^2x \partial_\mu n^a\partial_\mu n^a, \nonumber \\
&& n^an^a=1,\ a=1,2,3.
\end{eqnarray}
The spectrum of the model consists of a massive isospin
triplet of scalars.
The model displays spontaneus mass generation. It is
asymptotically free and
the coupling $g$ can be traded for the true parameter
of the model, $\Lambda$,
which is the scale characterizing the running coupling
in the $\overline{MS}$
scheme.
This is the well-known phenomenon of dimensional transmutation.
The ratio of the particle mass $m$ and the scale $\Lambda$
was calculated by Hasenfratz et al. \cite{hasi1,hasi2}
using Bethe ansatz techniques and the conjectured
exact S-matrix obtained
in \cite{zamzam}. This calculation also provides a
way of fixing the CDD
ambiguity by comparison with the $m/\Lambda$
ratio obtained from perturbation
theory.

We now make use of the idea which we call the
bootstrap fusion which was
used by Smirnov to calculate form-factors of the O(3)
nonlinear sigma model
from those of the SU(2) Thirring model. It is similar to the
description of
the bound states in the bootstrap approach but
in this case the limit is
singular (there are no actual bound states in the theory). We define

\begin{equation}
{\tilde Z}_I(\beta)=\lim\limits_{\epsilon\rightarrow 0}
AC_I^{ij}Z_i(\beta +
\pi i/2+i\epsilon)Z_j(\beta -\pi i/2-i\epsilon),
\end{equation}
where $A$ is a normal ordering constant
and $C_I^{ij}$ are the Clebsh-Gordan coefficients for the adjoint
representation of SU(2) in the product of
two fundamental representations.
(The capital letters denote adjoint indices,
taking the values $I=+,0,-$).
Due to proposition (ii) above the limit is
well-defined since the singularity
drops out. In
this way we obtain particle operators transforming
in the adjoint representation
of SU(2) or, the same, in the fundamental representation of O(3).
The nice
property, used by Smirnov, is the following:
if we now calculate what kind of
scattering relations are satisfied
by the ${\tilde Z}$'s what we find is the O(3)
sigma model S-matrix. The calculation proceeds by simple
commutations of the
$Z$'s and the result is

\begin{equation}
{\tilde Z}_I(\beta_1){\tilde Z}_J(\beta_2)=
{\tilde S}_{I,J}^{K,L}(\beta_{12})
{\tilde Z}_L(\beta_2){\tilde Z}_K(\beta_1),
\end{equation}
where ${\tilde S}$ denotes the S-matrix of the nonlinear
sigma model which is known to
be
\begin{eqnarray}
{\tilde S}_{I,J}^{K,L}(\beta )=&  &{ 1\over
(\beta +\pi i)(\beta -2\pi i)}\times\nonumber \\
&  &\left[\beta (\beta -\pi i)\delta_I^K\delta_J^L
-2\pi i(\beta -\pi i)\delta_I^L\delta_J^K
+2\pi i\beta\delta_{I,-J}\delta^{K,-L}\right] .
\end{eqnarray}

We use the spin-1/2 basis so the $C_I^{ij}$ are Clebsh-Gordan
coefficients of the quantum group SU(2)$\rm_{-1}$
for the spin-1 representation
in the product of two spin-1/2 representation.
Written out explicitely
${\tilde Z}_I(\beta )$ have the form
\begin{eqnarray}
{\tilde Z}_+(\beta)=&  &\lim\limits_{\epsilon\rightarrow 0}
AZ_+(\beta +
\pi i/2+i\epsilon)Z_+(\beta -\pi i/2-i\epsilon),\nonumber \\
{\tilde Z}_0(\beta)=&  &\lim\limits_{\epsilon\rightarrow 0}
A{1\over\sqrt{2}}
(Z_+(\beta + \pi i/2+i\epsilon)Z_-(\beta -\pi i/2-i\epsilon)
\nonumber \\
&  &-Z_-(\beta +\pi i/2+i\epsilon)
Z_+(\beta -\pi i/2-i\epsilon)),\nonumber \\
{\tilde Z}_-(\beta)=&  &\lim\limits_{\epsilon\rightarrow 0}
AZ_-(\beta +
\pi i/2+i\epsilon)Z_-(\beta -\pi i/2-i\epsilon).
\end{eqnarray}
We define the new vertex operator
\begin{eqnarray}
U (\beta )&  =\lim\limits_{\epsilon\rightarrow 0}\
:V(\beta +i\pi /2+i\epsilon ) V(\beta -i\pi /2-i\epsilon):
\nonumber \\
&  =\lim\limits_{\epsilon\rightarrow 0}A
V(\beta +i\pi /2+i\epsilon )V(\beta- i\pi /2-i\epsilon ).
\end{eqnarray}
{}From this equation the value of $A$ can be read off. However
this is just a normalization constant in the
pure quadratic scattering
relations and therefore can be dropped.
One can prove that the operator
$U$ can be expressed explicitely in terms of the free fields as
\begin{equation}
U(\beta )=\exp(i\bar\phi(\beta )),
\end{equation}
and satisfies the following relations
\begin{eqnarray}
&  U(\beta_1)U(\beta_2)
=&\bar g(\beta_{12}):U(\beta_1)U(\beta_2):, \nonumber \\
&  U(\beta_1)\bar V(\beta_2)
=&\bar g^{-1}(\beta_{12}):U(\beta_1)\bar V(\beta_2):.
\label{eq:normordering}\end{eqnarray}

Calculating the operator products and rearranging them
we arrive at the
following free field representation of the O(3) sigma model
Zamolodchikov-Faddeev operators
\begin{eqnarray}
&  {\tilde Z}_+(\beta )=&U(\beta ), \nonumber \\
&  {\tilde Z}_0(\beta )=&{i\over \sqrt{2}}
(\chi U (\beta )-U (\beta )\chi ), \nonumber \\
&  {\tilde Z}_-(\beta )=&{1\over 2}
(\chi\chi U (\beta )-2\chi U (\beta )\chi +
                        U (\beta )\chi\chi ).
\label{eq:o3zfopdef}\end{eqnarray}

These relations are easy to interpret. $U(\beta )$ is nothing else
than the highest weight in the multiplet on which the step operator
$\chi$ of the quantum group acts creating the lower weights.
The action
of $\chi$ (apart from normalization factors)
is the same as the adjoint
action of the SU(2)
(remember that this is the O(3) vector multiplet and
that the action of SU(2) and SU(2)$\rm _{-1}$
on integer spin representations
are the same as can be read off from the corresponding coproduct).

A remark on the derivation of the above
bosonization formulae is in order.
The derivation
of the first two expressions
is obvious but in the third case one has to consider two
screening charges. We consider first the vacuum-vacuum matrix element
of the relation.
As the labeling of the integration variables for the two
charges are irrelevant (since they are integrated out)
one may symmetrize the
integrand with respect to the
rapidity argument of the two screening charges.
With this trick one can prove the equality of the integrands.
Then one checks
that the contours can be deformed
to prove the equality in the manner described
in \cite{freefield}.
The last thing is that since
eq. (\ref{eq:o3zfopdef}) is an operator relation one
must prove it for all matrix elements.
This can be achieved by writing down
the relation between general multiparticle states.
As a straightforward
calculation shows, the proof reduces to
using the same formulae as for
vacuum matrix elements.

The last thing to check is
the residue equation for the ZF operators of
the O(3) sigma model. Namely, in order to solve Smirnov's form-factor
equations one has to reproduce the kinematical poles at the rapidity
arguments displaced by $i\pi$. This means that we need
\begin{equation}
{\rm res}_{\beta_1=\beta_2+i\pi}
{\tilde Z}_I(\beta_1){\tilde Z}_J(\beta_2)=-iC_{IJ}
\end{equation}
However for the operators defined in
(\ref{eq:o3zfopdef}) this relation is not
fulfilled. The problem is that if we write
out the explicit representation of the O(3)
ZF operators with the help
of the bootstrap fusion in terms of the
SU(2) Thirring model ZF operators,
we can see that we do not obtain the correct pole structure.
We return to this question later
when we discuss an explicit example of
the solution for the bootstrap equations.

\section{The form-factor equations and their solution}

We call form-factor the matrix element
of a local field operator between
the vacuum and some $n$-particle
in-state $\vert\beta_1,i_1,\dots ,\beta_n,i_n
\rangle$ ($i_1,\dots i_n$ denote the
possible internal quantum numbers of
the corresponding particle):
\begin{equation}
F_{i_1\dots i_n}(\beta_1,\dots\beta_n)=
\langle vac\vert O(0)\vert\beta_1,i_1,
\dots ,\beta_n,i_n\rangle .
\end{equation}
Knowledge of these objects permits the reconstruction
of the matrix elements
of local operators between any many-particle states in the theory by
analytic continuation.
Also if the theory satisfies asymptotic completeness
with this set of many-particle states, then an infinite series
can be written down
for any correlation function.
Here we will be concerned by finding an integral
representation for the form-factors.
They are constrained by the Smirnov
axioms which may be taken as defining
the space of local operators. The
axioms are the following:
\bigskip

1. The functions $F$ are analytic in the rapidity differences
$\beta_{ij}=
\beta_i-\beta_j$ inside the strip $0<{\rm Im}\beta <2\pi$ except for
simple poles.
The physical matrix elements are the values of the $F$'s at
real rapidity arguments ordered as $\beta_1<\beta_2<\dots <\beta_n$.

2. The form-factor of a local operator of Lorentz spin $s$ satisfies
\begin{equation}
F_{i_1,\dots ,i_n}(\beta_1+\Lambda ,\dots ,\beta_n+\Lambda )=
\exp (s\Lambda )F_{i_1,\dots ,i_n}(\beta_1 ,\dots ,\beta_n ).
\end{equation}
This means that the
form-factor of a spin $s=0$ operator depends only on
the rapidity differences.

3. Form factors satisfy Watson's symmetry property:
\begin{eqnarray}
&  F_{i_1,\dots ,i_{j},i_{j+1},\dots ,i_n}
(\beta_1,\dots ,\beta_j,\beta_{j+1},
\dots ,\beta_n)=\nonumber \\ &  S_{i_{j}i_{j+1}}^{k_{j+1}k_{j}}
F_{i_1,\dots ,k_{j+1},k_{j},\dots ,i_n}
(\beta_1,\dots ,\beta_{j+1},\beta_{j},
\dots ,\beta_n).
\end{eqnarray}

4. Form factors satisfy the cyclic property
\begin{equation}
F_{i_1,\dots ,i_n}(\beta_1,\dots ,\beta_n+2\pi i)=
F_{i_n,i_1,\dots ,i_{n-1}}(\beta_n,\beta_1,\dots ,\beta_{n-1}).
\end{equation}

5. Form factors have kinematical singularities which are simple poles
at points where two of the rapidity arguments are displaced by $i\pi$.
The residue is given by the following formula
\begin{eqnarray}
&  {\rm res}_{\beta_i=\beta_n+i\pi}
F_{i_1,\dots ,i_n}(\beta_1,\dots ,\beta_n)=
-iC_{i_n,i_j^,}
F_{i_1^{,},\dots ,\hat i_j^,\dots ,i_{n-1}^{,}}(\beta_1,\dots ,
\hat\beta_j,\dots ,\beta_n)\times\nonumber \\
&  [\delta_{i_1}^{i_1^{,}}\dots\delta_{i_{j-1}}^{i_{j-1}^{,}}
S_{i_{n-1}k_1}^{i_{n-1}^{,}i_j^{,}}(\beta_{n-1}-\beta_j)
S_{i_{n-2}k_2}^{i_{n-2}^{,}k_1}(\beta_{n-2}-\beta_j)\dots
S_{i_{j+1}i_j}^{i_{j+1}^{,}k_{n-j-2}}(\beta_{j+1}-\beta_j)\nonumber \\
&  -S_{k_1,i_1}^{i_{j}^{,}i_1^{,}}(\beta_j-\beta_1)\dots
S_{k_{j-2}i_{j-2}}^{k_{j-3}i_{j-2}^{,}}(\beta_j-\beta_{j-2})
S_{i_ji_{j-1}}^{k_{j-2}i_{j-1}^{,}}(\beta_j-\beta_{j-1})
\delta_{i_{j+1}}^{i_{j+1}^{,}}
\dots\delta_{i_{n-1}}^{i_{n-1}^{,}} ],\nonumber \\
\label{eq:formfactoraxioms}\end{eqnarray}
where the hat denotes the omission of
the corresponding rapidity argument and
internal index.

Since in the O(3) sigma model
we assume no bound states, there are no more
poles. If one has bound states then it is necessary
to add a new axiom stating
the position and the residue of the bound state poles.

Smirnov proved the following important theorem:
if two operators are defined
by matrix elements satisfying the axioms (1-5)
then they are mutually local.

This is called the locality theorem
and it has a profound implication; namely
we can take the space of operators defined by the
solutions of the axioms
(\ref{eq:formfactoraxioms}) to generate
the local operator algebra of the model.

Now we proceed to give a formula for the solutions of the equations
(\ref{eq:formfactoraxioms}) which was
first written down by Lukyanov. Before doing it
we comment on the omission of
the conjugate ZF operators ${\tilde Z}^\dagger(\beta )$.
They are certainly needed in
order to build up the space of asymptotic states.
However one can define a representation of the ZF algebra
as follows \cite{freefield}.
We take a ground state
$\vert 0\rangle$
(this is not the vacuum state of the theory!) and define a
module of the ZF algebra by acting with the ${\tilde Z}'s$ on it.
Specifically we
define a representation $\pi_Z$ on the
space ${\cal H}_Z$ which is given by
\begin{equation}
{\cal H}_Z=
{\rm span}\{ Z_{i_1}(\beta_1)\dots Z_{i_n}(\beta_n)\vert 0\rangle \}.
\end{equation}
This representation can be interpreted as
doing quantum field theory in Rindler space and using
angular quantization. This was explained in \cite{jostfunc}.

Having the representation discussed above one looks for the following
structures:

1. The operator of Lorentz boosts $K$ satisfying
\begin{equation}
Z_i(\beta+\Lambda )=\exp (-\Lambda K)Z_i(\beta)\exp (\Lambda K).
\end{equation}

2. A map $O\rightarrow L(O)$
from the space of local operators to the algebra
of endomorphisms of $\pi_Z$ satisfying the following two conditions
\begin{eqnarray}
  L(O)Z_i(\beta )& =& Z_i(\beta )L(O), \nonumber \\
 \exp ( \Lambda K)L(O)\exp (-\Lambda K)& =
&\exp (\Lambda s(O))L(O).
\end{eqnarray}
where $s(O)$ denotes the Lorentz spin of the operator $O$.

If one has these objects then, as shown by Lukyanov \cite{freefield},
the functions
\begin{equation}
F_{i_1,\dots i_n}(\beta_1 ,\dots ,\beta_n)=
Tr_{\pi_Z} [\exp (2\pi iK)L(O)
Z_{i_n}(\beta_n)\dots Z_{i_1}(\beta_1)],
\end{equation}
are solutions of the form-factor axioms.

We now have to give the definition of the dual operators
which give us the
map $L$ representing the local operators of the model.
We introduce the
scalar field $\phi^\prime (\alpha )$
satisfying the same relations as $\phi$
together with its associated field $\bar\phi^\prime (\alpha )$.
(We give a
closer definition of these fields shortly
when we introduce an ultraviolet
regularization for the free field system.) With the help
of these new fields we define the dual vertex operators as follows:
\begin{eqnarray}
&  \Lambda_-(\alpha )=&{U^\prime}(\alpha ), \nonumber \\
&  \Lambda_0(\alpha )=&{i\over\sqrt{2}}[{\chi^\prime}{U^\prime}(\alpha )
-{U^\prime}(\alpha ){\chi^\prime} ], \nonumber \\
&  \Lambda_+(\alpha )=&{1\over 2}[{\chi^\prime}^2{U^\prime}
(\alpha )-2{\chi^\prime}
{U^\prime}(\alpha ){\chi^\prime}+{U^\prime}(\alpha ){\chi^\prime}^2].
\end{eqnarray}
The dual vertex operators ${U^\prime}$ and the $\bar V^\prime$ required for
${\chi^\prime}$, satisfy similar relations as the unprimed ones, but with
the function $\bar g$ changed to
\begin{equation}
\bar g^\prime(\alpha )=-k^2{\alpha (\alpha -i\pi )\over 4\pi^2}.
\end{equation}
The normalization of ${\chi^\prime}$
which is the analog of the parameter $\eta$
for $\chi$ is chosen to equal $(i\pi )^{1/2}$.

As can be seen from the above definition,
${\chi^\prime}$ plays the role of the
other step operator for the quantum group.
With the operator $P$ to be
introduced later, $\chi$ and ${\chi^\prime}$ satisfy
all the algebraic relations
of $SU(2)_{-1}$ and obey the correct
coproduct rule. The proof of this
statement can be found in \cite{freefield}.

One can also calculate all the
normal ordering relations involving the
primed and unprimed operators merely by using the relation
\begin{equation}
\bar V^\prime(\delta )V(\beta )=\rho\bar\rho^\prime u(\beta -\delta)
:\bar V^\prime(\delta )V(\beta ):,
\end{equation}
where the function $u$ is given by
\begin{equation}
u(\beta )=k{i\beta\over 2\pi}.
\end{equation}

Now one writes down the functions
\begin{eqnarray}
&&  {\cal F}^{i_1,\dots i_k}_{j_1,\dots j_l}
(\alpha_1,\dots \alpha_k\vert
\beta_1,\dots \beta_n)\nonumber \\ &&  =
Tr_{\pi_Z} [\exp (2\pi iK)\Lambda_{i_k}(\alpha_k)
\dots\Lambda_{i_1}(\alpha_1){\tilde Z}_{j_l}(\beta_l)
\dots{\tilde Z}_{j_1}(\beta_1)],
\end{eqnarray}
then these functions have to be expanded as
\begin{eqnarray}
&&  {\cal F}^{i_1,\dots i_k}_{j_1,\dots j_l}
(\alpha_1,\dots \alpha_k\vert
\beta_1,\dots \beta_n)\nonumber \\ &&
= \sum\limits_{\{s_j\}} F^{\prime i_1,\dots i_k}
(\alpha_1,\dots \alpha_k\vert\{s_j\})F_{j_1,\dots j_l}(\{s_j\}\vert
\beta_1,\dots \beta_n)\times\nonumber \\ &&
\exp (s_1\alpha_1)\dots\exp (s_k\alpha_k),
\end{eqnarray}
where the functions $F$ are the required solutions of the form-factor
axioms. This means that the functions
$\cal F$ play the role of generating
functions for the form-factors of local operators.

The actual calculation of the matrix elements and the traces proceeds
through the procedure described in \cite{freefield}.
The free field construction needs a regularization;
we choose an ultraviolet
cut-off by taking the rapidity interval finite
\begin{equation}
-{\pi\over\epsilon}<\beta <{\pi\over\epsilon}.
\end{equation}
Then we introduce the mode expansion of the free field $\phi$ as:
\begin{equation}
\phi_\epsilon(\beta )=
{1\over\sqrt{2}}(Q-\epsilon\beta P)+\sum\limits_{k\neq 0}
{a_k\over i\sinh (\pi k\epsilon )}\exp (ik\epsilon\beta),
\end{equation}
where the oscillator modes satisfy the commutation relation
\begin{equation}
[a_k,a_l]={\sinh{\pi k\epsilon\over 2}\sinh\pi k\epsilon\over k}
\exp{\pi\vert k\vert\epsilon\over 2}\delta_{k,-l},
\end{equation}
and the zero modes satisfy the canonical commutation relation
\begin{equation}
[P,Q]=-i.
\end{equation}
The two-point function and the
commutation relations for this field can
be calculated and it can be seen that in the
limit $\epsilon\rightarrow 0$
we recover the relations satisfyed by the field
$\phi$ (see eq. (\ref{eq:phirelations})). The dual field is given by
\begin{equation}
\phi^\prime_\epsilon(\alpha )=-{1\over\sqrt{2}}(Q-\epsilon\alpha P)+
\sum\limits_{k\neq 0}
{a^\prime_k\over i\sinh (\pi k\epsilon )}\exp (ik\epsilon\alpha),
\end{equation}
with the oscillators $a_k^\prime$ defined by the relation
\begin{equation}
a_k^\prime\exp\left({\pi\vert k\vert\epsilon\over 4}\right)=
a_k\exp \left( -{\pi\vert k\vert\epsilon\over 4}\right)
\end{equation}
The space where these operators live is defined to be
\begin{equation}
\pi^\epsilon_Z = \bigoplus\limits_{k\in \rm Z} F_{k/\sqrt{2}},
\end{equation}
where $F_p$ denotes the Fock space
built up with the help of the creation
operators (as usual, the oscillator modes with
negativ indices) from the
ground state $\vert p\rangle$ which satisfies:
\begin{equation}
P\vert p\rangle=p\vert p\rangle.
\end{equation}
The operator $K$, of the Lorentz boost is given by
\begin{equation}
K_\epsilon=i\epsilon H - i{\sqrt{2}\over4}\epsilon P,
\end{equation}
where
\begin{equation}
H={P^2\over2}+\sum\limits_{k=1}^\infty
{k^2\over\sinh{\pi k\epsilon\over 2}
\sinh\pi k\epsilon}a^\prime_{-k}a_k.
\end{equation}
Formally we can write
$\pi_Z=\lim\limits_{\epsilon\rightarrow 0}\pi^\epsilon_Z$.
This should be understood in the following sense:
we make all calculations
with the regularized operators and in the end take the limit
$\epsilon\rightarrow 0$.

\section{Calculation of the traces and the problem of the integration}

We should calculate the traces required for obtainig the integrand of
the representation for the form-factors.
In \cite{freefield} it was proposed
that one should use the technique of
Clavelli and Shapiro which they invented
for the calculation of string scattering amplitudes.
The prescription is to
introduce another copy of the oscillators $a_k$
which commute with $a_k$
and satisfies the same commutation among themselves.
Then we make the
following definitions:
\begin{eqnarray}
&  \tilde a_k=
&{a_k\over{1-\exp (-2\pi k\epsilon )}}+b_{-k}\ (k>0),\nonumber \\
&  \tilde a_k=
& a_k+{b_{-k}\over{1-\exp (-2\pi k\epsilon )}}\ (k<0).
\end{eqnarray}
For an operator $O({a_k})$ on the Fock space $F(a)$
we introduce the operator
$\tilde O=O({\tilde a_k})$ by substituting $\tilde a_k$ for $a_k$.
Then we
can calculate the regularized trace
over the non-zero modes as ($Tr_F$ denotes
the trace over the Fock module of non-zero modes):
\begin{equation}
Tr_F[\exp (2\pi iK_\epsilon )O]=
{\langle 0\vert \tilde O\vert 0\rangle\over
\prod\limits_{k=1}^{\infty}(1-\exp (-2\pi m\epsilon ))},
\end{equation}
and then take the limit $\epsilon\rightarrow 0$.
The state $\vert 0\rangle$ is
nothing else but the ground state in the
product Fock module $F[a]\bigotimes
F[b]$. The trace over the zero modes just gives the $SU(2)$
superselection rules for
the matrix elements.
Using the results in \cite{freefield} we get for the
general trace
\begin{eqnarray}
Tr_{\pi_Z}[\exp (2\pi iK)U^\prime (\alpha_k)\dots U^\prime (\alpha_1)
\bar V^\prime (\delta_p)\dots\bar V^\prime (\delta_1)
U (\beta_n)\dots U (\beta_1)\bar V(\gamma_r)\dots\bar V(\gamma_1)]\
nonumber \\
\end{eqnarray}
the following result:
\begin{eqnarray}
&&{\cal C}_1^{-n} {\cal C}_2^{n+r\over 2}
{\cal C}_2^{\prime {p+k\over 2}} 2^{-r-n}
i^{n+p+k}\eta^r\eta^{\prime -p-k}
\delta_{2n-2r+2p-2k,0}
\prod\limits_{i<j}\bar G(\beta_i-\beta_j)
\prod\limits_{i<j}\bar G(\gamma_i-\gamma_j)\times\nonumber \\
&&\prod\limits_{i,j}\bar G^{-1}(\gamma_i-\beta_j)
\prod\limits_{i<j}\bar G^\prime(\alpha_i-\alpha_j)
\prod\limits_{i<j}\bar G^\prime(\delta_i-\delta_j)
\prod\limits_{i,j}\bar G^{\prime -1}(\delta_i-\alpha_j)
\prod\limits_{i,j}\bar H(\beta_i-\alpha_j)\times\nonumber \\
&&\prod\limits_{i,j}\bar H^{-1}(\beta_i-\delta_j)
\prod\limits_{i,j}\bar H(\gamma_i-\delta_j)
\prod\limits_{i,j}\bar H^{-1}(\gamma_i-\alpha_j).
\end{eqnarray}

Here we defined the new functions as:
\begin{eqnarray}
\bar G(\beta )& =& -{ {\cal C}_2\over  4}
(\beta+i\pi )\sinh\beta, \nonumber \\
\bar G^\prime (\alpha )& =& -{\cal C}_2^\prime
{ \sinh\alpha\over \alpha +i\pi}, \nonumber \\
\bar H(\alpha )& =& -{ 2\over \cosh\alpha},
\end{eqnarray}
and the constants are given by:
\begin{eqnarray}
&& {\cal C}_1=\exp\left[ -\int\limits_0^\infty { dt\over t}
{\displaystyle \sinh^2 {t\over 2}\exp (-t)
\over \sinh 2t \cosh t}\right],\nonumber \\
&& {\cal C}_2={\cal C}_2^{\prime -1}
={\displaystyle \Gamma^4({1\over4})\over\displaystyle 4\pi^3}.
\end{eqnarray}
(see \cite{freefield}).
The trace over the zero modes however contains an
infinite constant which we have set equal to unity.

We also have to specify the integration
over the variables of the screening
charges. We face the following problem:
when we take the integration contour
as given in the SU(2) Thirring model (see in \cite{freefield})
and try to take the limit which gives
the O(3) sigma model, new double poles arise
in the integration over the
$\gamma_i$ variables whose positions depend
on the $\beta_j$ variables.
This phenomenon was already noticed by
Smirnov \cite{smirnov4}.
The contour gets pinched
by the two poles approaching each other and the integrals diverge.
This means
that the limit to the O(3) model is singular.
To deal with the integral we
introduce the prescription that one
should take the coefficient of the most
divergent term in the integral.
This is equivalent to the procedure that
Smirnov took.
The other integrals cause no problems: the contours are specified
as earlier for the calculation of vacuum matrix elements
but they must lie
within the strip $-i\pi <\gamma_i,\delta_j<i\pi$.
In the following section
we show the calculation for the simplest nontrivial case.

\section{An explicit example}

What we will calculate using the method
outlined above is the two-particle
form-factor for the O(3) current.
This was known long before but what we
obtain is the generating function for the two-particle form-factors
of a sequence
of isovector operators with different Lorentz spin.
The current is just the
lowest element of the sequence.

The two-particle form-factor of the current is the matrix element:
\begin{equation}
f^{ABC}_\mu(\beta_1,\beta_2)=\langle 0\vert j^A_\mu (0)
\vert \beta_1,B,\beta_2,C\rangle.
\end{equation}
The tensor structure of the form-factor
is given by the Levi-Civita symbol
$\epsilon^{ABC}$.
Since this matrix element has only one
independent component it is enough
to calculate only one function. This is given by the following:
\begin{equation}
{\cal F}(\alpha ,\beta_1,\beta_2)=
Tr_{\pi_Z} [\exp (2\pi iK)Z^\prime_- (\alpha)
Z_+(\beta_1)Z_0(\beta_2)].
\end{equation}
This is the simplest matrix element
allowed by the superselection rules
which possesses the required isospin properties.
The trace we have to calculate
is:
\begin{equation}
Tr_{\pi_Z} [ \exp (2\pi iK) U^\prime (\alpha )U (\beta_1)
(\bar V(\gamma )U (\beta_2) -U (\beta_2) \bar V(\gamma ))],
\end{equation}
which gives the following result
(up to a constant multiplier, which is unknown since
the trace of the zero modes, as noted above,
contains an infinite constant):
\begin{eqnarray}
&  {\displaystyle 1\over\displaystyle
\cosh (\beta_1-\alpha )\cosh (\beta_2-\alpha )}
(\beta_{21}+i\pi )
\sinh(\beta_{21})\nonumber \\
&  \Biggl[ {\displaystyle\cosh (\gamma -\alpha )\over\displaystyle
(\gamma -\beta_1+i\pi ) \sinh (\gamma -\beta_1)
(\beta_2-\gamma +i\pi)\sinh (\beta_2-\gamma )} -\nonumber \\
&  {\displaystyle\cosh (\gamma -\alpha )
\over\displaystyle (\gamma -\beta_1+i\pi )
\sinh (\gamma -\beta_1) (\gamma -\beta_2+i\pi)
\sinh (\gamma-\beta_2 )} \Biggr].
\end{eqnarray}
The integration over the arguments of the screening charges gives:
\begin{eqnarray}
&  {\displaystyle 1\over\displaystyle \cosh (\beta_1-\alpha )
\cosh (\beta_2-\alpha )}
(\beta_{21}+i\pi ) \sinh(\beta_{21})\nonumber \\
&  \Biggl[ \displaystyle \int\limits_{C_1}{\displaystyle
d\gamma\over \displaystyle 2\pi}
{\displaystyle\cosh (\gamma -\alpha )\over \displaystyle
(\gamma -\beta_1+i\pi )\sinh (\gamma -\beta_1)
(\beta_2-\gamma +i\pi)\sinh (\beta_2-\gamma )} - \nonumber \\
&  \displaystyle \int\limits_{C_2}
{\displaystyle d\gamma\over \displaystyle 2\pi}
{\displaystyle \cosh (\gamma -\alpha )\over \displaystyle
(\gamma -\beta_1+i\pi )\sinh (\gamma -\beta_1)
(\gamma -\beta_2+i\pi)\sinh (\gamma-\beta_2 )} \Biggr],
\end{eqnarray}
where the contours run on the
complex $\gamma$ plane from $\Im \gamma=-\infty$
to $\Im \gamma=\infty$ and pass through the double poles which are
$\gamma =\beta_1-i\pi$
and $\gamma =\beta_2+i\pi$ in the case of $C_1$ and
$\gamma =\beta_1-i\pi$ and
$\gamma =\beta_2-i\pi$ in the case of $C_2$. As we
said before we interpret this integral
as the limit of two coinciding poles
giving a double pole and calculate
the coefficient of the most singular term.
This is a procedure that can be easily seen
to be consistent with the
form-factor axioms 2-4 (see section 4)
but it does not yield the correct
analytic structure. It was already noted
when we discussed the singularities
of the operator product of the ZF operators (see end of section
\ref{ZFsing}) that
there will be problems with kinematical singularities.
Performing this calculation yields the following:
\begin{equation}
{\displaystyle (\beta_{21}+i\pi )\sinh(\beta_{21})\over
\displaystyle\cosh (\beta_1-\alpha )\cosh (\beta_2-\alpha )} \left[
{\displaystyle\cosh (\beta_2-\alpha )
\over \displaystyle(\beta_{21}+2i\pi )\sinh
(\beta_{21})}- {\displaystyle\cosh (\beta_1-\alpha )\over
\displaystyle\beta_{21}\sinh (\beta_{21})} \right],
\end{equation}
which simplifies to
\begin{equation}
{2i\pi (\beta_{21}+i\pi )\over (\beta_{21}+2i\pi )\beta_{21} }
\left[{1\over \cosh(\beta_1-\alpha )}
+{1\over \cosh(\beta_2-\alpha )}\right].
\end{equation}
Here we see the problem with the analytic structure:
the function above
does not have the correct kinematical pole at
$\beta_2=\beta_1+i\pi$ but it
does have poles at $\beta_2=\beta_1$
and $\beta_2=\beta_1+2i\pi$. This can be
dealt with by introducing a
factor which does not destroy the symmetry
properties of the function but
restores the correct analiticity. The factor
must be a function of rapidity difference only,
must satisfy perodicity and
symmetry and it must have the
correct analitical properties. A function with
these properties is
\begin{equation}
f(\beta_{21})= \tanh^2{\beta_{21}\over 2}.
\end{equation}
In the case of general $2n$-particle form-factor
the correct procedure is to
introduce such factors for each pairing of
particle rapidities. This is true since
the analytic behaviour in all two-particle rapidity differences
is the same ( because the symmetric
construction of the form-factor solutions).
This appears to be a general prescription for curing the bad analytic
behaviour (see \cite{smirnov4}).
So the final generating function is:
\begin{equation}
{2i\pi (\beta_{21}+i\pi )\over (\beta_{21}+2i\pi )\beta_{21} }
\left[{1\over \cosh(\beta_1-\alpha )}
+{1\over \cosh(\beta_2-\alpha )}\right]
\tanh^2{\beta_{21}\over 2}.
\end{equation}
Now expanding the generating function around $\exp(\alpha )=0$ and
$\exp (-\alpha )=0$, respectively, we find the first coefficient to be
\begin{equation}
{2i\pi (\beta_{21}+i\pi )\over (\beta_{21}+2i\pi )\beta_{21} }
[\exp (\beta_1)+\exp (\beta_2)]
\tanh^2{\beta_{21}\over 2},
\end{equation}
and
\begin{equation}
{-2i\pi (\beta_{21}+i\pi )\over (\beta_{21}+2i\pi )\beta_{21} }
[\exp (-\beta_1)+\exp (-\beta_2)]
\tanh^2{\beta_{21}\over 2}.
\end{equation}
These are the two-particle form-factors
of two operators with isospin $1$ and
Lorentz spin $\pm 1$, respectively,
which can be identified as the two
light-cone components of the O(3) current.
Expanding the generating function
to higher orders the form-factors of infinitely
many operators can be found.

In the literature one can find the result
for this form-factor \cite{balog}
which we quote for comparison:
\begin{eqnarray}
&& \langle 0\vert j_\omega^A (0)\vert\beta_1,B,\beta_2,C\rangle
=-{\displaystyle i\pi
\over\displaystyle 8}\epsilon^{ABC}
{\displaystyle\beta_{12}-i\pi\over\displaystyle\beta_{12}
(2i\pi-\beta_{12})}
\tanh^2{\displaystyle\beta_{12}\over\displaystyle 2}\nonumber \\
&& (-m\omega )
\left( \exp (-\omega\beta_1)+\exp (-\omega\beta_2)\right).
\end{eqnarray}
This agrees with the result derived
here up to a normalization factor. This
factor
can be fixed if one requires that
the charge obtained from the O(3) current
should take the correct value on
the many-particle states. Here $\omega=\pm$
denotes the Lorentz components of the current.

\section{Conclusions}

We would like now to summarize the
results briefly. We succeeded in writing
down a free field representation for the
O(3) ZF algebra and obtained a
representation for the solutions of the
form-factor equation. We saw that the
representation obtained by the fusion
procedure is not completely satisfactory
since a hand-made, although simple correction,
is needed to achieve the
necessary
analytical structure of the form-factors.
First we would like to comment on
the origin and the possible way of correcting for this problem.

What lies behind all the construction is that
one divides the S-matrix into
two factors: a scalar function, which is nothing
else but the scattering
amplitude in the highest isospin channel
(see the function $S_0(\beta )$ in the
case of SU(2) Thirring model) and a tensor part,
which carries all of the
isospin structure of the S-matrix.
Then we solve a Riemann-Hilbert problem
and find the function which obeys the equation
\begin{equation}
g(-\beta )=S_0(\beta )g(\beta ).
\end{equation}
With the help of this function a free field
can be defined which has $g$
as the two-point function.
But the solution of the above equation is not unique;
in fact, there is an ambiguity up to an even function as can be seen
easily. The solution must be fixed by requiring
some analiticity conditions,
which in turn fixes the singularities in the
operator products and the
pole structure of the form-factors.
This suggests that the factorization
of the O(3) S-matrix obtained from the bootstrap
is not the correct one
and this necessitates the introduction of the
extra $\tanh^2$ factors
in the form-factor expression.
A possible way out would be to find a new
factorization of the O(3) S-matrix,
which gives a new free field representation
satisfying the required analytic properties.
We hope to investigate this
possibility further.

The above discussion also shows that there may be a generalization of
this free field construction
to at least a class of further integrable
models; it seems to be important to clear up
whether such a generalization
exists and try to construct it.
Another relevant question is whether the
family of solutions that can be obtained
from the free field representation
is complete. We mean by this completeness the
question of whether the
operators defined by all of these solutions form a
closed algebra of local
operators. If the answer is yes then the constructed
set of operators should be
taken as the definition of the theory in the
sense of algebraic quantum
field theory and then the question of the dynamics can be thoroughly
investigated. Namely one can try to
show that the fields constructed in
this way satisfy the equations of motion.
For example, the constraint
equation for the O(3) field ($n^an^a=1$)
can be proved if it can be shown that
there exists no Lorentz and isoscalar operator
with canonical dimension zero
other than the unit operator in the model. This can be checked by
showing that the two and
higher particle form-factors of such an operator
necessarily vanish. There is some evidence for that (the two-particle
form-factor vanishes and within
the frame of a polynomial ansatz this is
also true for the four-particle one \cite{balogpriv}).
The other field
equation is just the conservation
of the O(3) current, which is satisfied
for all many-particle form-factors of
the current (see \cite{smirnov4}).
However one should also prove the
relation between the field $n^a$ and the
current $j^a_\mu$ which looks
\begin{equation}
j^a_\mu =\epsilon^{abc}n^b\partial_\mu n^c
\end{equation}
A confirmation for this relation would be to
show that there is no SU(2)
singlet conserved current in the operator algebra
since that would mean
$j^an^a=0$.

The discussion of the form-factors is
relevant to the question of correlation
functions as well,
since the correlation functions can be obtained in the
model from the form-factors in terms of infinite series expansions.
Up to now, however,
there is no nice way to handle such series expansions and
deciding whether
they are convergent and how fast they converge. However
some recent studies \cite{balog}
show that they can be compared to the
perturbation theory results and
they show a reasonable agreement also with
lattice simulation results.
With a better understanding of the properties of
form-factors one may get closer
to the solution of these problems as well.

\acknowledgements
The authors would like to
thank to L. Palla for extensive discussions.
This work was partially
supported by the Hungarian National Science and
Research Foundation (Grant No. 2177).
\appendix

\section{A direct proof for the ZF relations}

We would like to present
a method for proving the ZF scattering relations
for the O(3) model.
These relations are the analogues of (\ref{eq:scattrel}) with
the O(3) S-matrix
instead of the one for the SU(2) Thirring model. The
reasoning given above used the fusion procedure.
There is also a direct way of
demonstrating the O(3) ZF relations without relying on the connection
between the SU(2) Thirring model and the O(3) sigma model.

The relation we want to prove is

\begin{equation}
{\tilde Z}_I(\beta_1){\tilde Z}_J(\beta_2)=
{\tilde S}_{I,J}^{K,L}(\beta_{12})
{\tilde Z}_L(\beta_2){\tilde Z}_K(\beta_1).
\end{equation}

This means nine equations
to prove because the indices $I,J,K,L$ take on
3 distinct values.
We will classify these equations by the number of the
screening charges appearing in them.
The only relation without screening
charges is:
\begin{equation}
{\tilde Z}_+(\beta_1){\tilde Z}_+(\beta_2)=
{\tilde S}_{+,+}^{+,+}(\beta_{12})
{\tilde Z}_+(\beta_2){\tilde Z}_+(\beta_1).
\end{equation}
Substituting here the
expression (\ref{eq:o3zfopdef}) for ${\tilde Z}_+$ and normal ordering
the two side of the equation
with the help of eq. (\ref{eq:normordering}) one
can see that the above relation is trivially satisfied.

There are two relations with one screening charge.
We discuss one of them since
the other one is similarly treated. We should prove that
\begin{equation}
{\tilde Z}_+(\beta_1){\tilde Z}_0(\beta_2)=
{\tilde S}_{+,0}^{0,+}(\beta_{12})
{\tilde Z}_+(\beta_2){\tilde Z}_0(\beta_1)+
{\tilde S}_{+,0}^{+,0}(\beta_{12})
{\tilde Z}_0(\beta_2){\tilde Z}_+(\beta_1).
\end{equation}
We substitute here eq. (\ref{eq:o3zfopdef}) to get
\begin{eqnarray}
U(\beta_1)(\chi U (\beta_2)-U (\beta_2)\chi )=
{\tilde S}_{+,0}^{0,+}(\beta_{12})
U (\beta_2)(\chi U (\beta_1)-U (\beta_1)\chi )+\nonumber \\
{\tilde S}_{+,0}^{+,0}(\beta_{12})
(\chi U (\beta_2)-U (\beta_2)\chi )U (\beta_1).
\end{eqnarray}
The screening charge contains contour integration. We can write out
the integrands and
after normal ordering according to eq. (\ref{eq:normordering})
we get the following for the equality of the integrands:
\begin{eqnarray}
{\bar g} (\beta_2-\beta_1){\bar g}^{-1}
(\gamma -\beta_1)({\bar g}^{-1} (\beta_2-\gamma )-
{\bar g}^{-1}(\gamma -\beta_2))=\nonumber \\
{\tilde S}_{+,0}^{0,+}(\beta_{12})
{\bar g} (\beta_1-\beta_2){\bar g}^{-1}
(\gamma -\beta_2)({\bar g}^{-1} (\beta_1-\gamma )-
{\bar g}^{-1}(\gamma -\beta_1))+\nonumber \\
{\tilde S}_{+,0}^{+,0}(\beta_{12})
{\bar g} (\beta_1-\beta_2){\bar g}^{-1}
(\beta_1-\gamma )({\bar g}^{-1} (\beta_2-\gamma )-
{\bar g}^{-1}(\gamma -\beta_2)),
\end{eqnarray}
which is a simple
algebraic identity that can be proven by a bit of calculation.
The variable $\gamma$
is the integration varible for the screening charge
$\chi$. Then one can integrate over $\gamma$.
Using the rules for the contours
given in section \ref{contours} one
can draw the different contours on the complex $\gamma$
plane and show that they can be deformed
into each other without encountering
any poles. This completes the
proof of this particular case. The other
case with
one screening charge corresponding to the scattering relation
${\tilde Z}_0{\tilde Z}_+\rightarrow
{\tilde Z}_0{\tilde Z}_++{\tilde Z}_+{\tilde Z}_0$
( in a selfexplanatory notation
where $0,+,-$ denote the SU(2) quantum number of the
corresponding ZF operators) can be proven similarly.

The case of two screening charges
is much more involved so we just sketch
the method.
There are 3
such identities: the ${\tilde Z}_0{\tilde Z}_0
\rightarrow {\tilde Z}_0{\tilde Z}_0+{\tilde Z}_+{\tilde Z}_-
+{\tilde Z}_-{\tilde Z}_+$,
the ${\tilde Z}_+{\tilde Z}_-
\rightarrow {\tilde Z}_0{\tilde Z}_0+
{\tilde Z}_+{\tilde Z}_-+{\tilde Z}_-{\tilde Z}_+$ and the
${\tilde Z}_-{\tilde Z}_+\rightarrow
{\tilde Z}_0{\tilde Z}_0+{\tilde Z}_+{\tilde Z}_-
+{\tilde Z}_-{\tilde Z}_+$
case.
Since we have two integration variables we can symmetrize in them
since the
integral is independent of which is named $\gamma_1$ and which is
named $\gamma_2$.
We verified the identities by doing calculations with
Maple. The
case of three and four screening charges is in principle also
similar,
but quite tedious so it is better to argue with the help of the
bootstrap fusion idea.

\end{document}